**From Local Atomic Motifs to Thermodynamic State: An Interpretable Physics-Informed Framework for Cu–Zr Metallic Glasses**

Prashil S. Joshi[1*]

(1) Prashil S. Joshi[1] (***Corresponding author**)

E-mail: prashiljoshi@iisc.ac.in

Affiliation(s): [1]Department of Materials Engineering, Indian Institute of Science, Bengaluru, Karnataka - 560012, India.

ORCID: orcid.org/0009-0009-6727-5483

**Abstract:**

Machine-learning models that relate local atomic structure to the thermodynamic state of metallic glasses typically assess physical consistency after training rather than enforcing it during learning. Here, we develop a multi-task physics-informed neural network (PINN) that predicts temperature directly from Voronoi-motif population histograms while incorporating autograd-derived gradient constraints representing physically motivated relationships between structural motifs, quench rate, and temperature. The model simultaneously classifies each configuration as liquid, supercooled/transition, or glass through an auxiliary classification head. The classification task achieves 96.6% test accuracy with a macro-F1 score of 0.95, while the regression head yields a mean absolute error of 23.4 K on a trajectory-disjoint held-out test set. Five-fold trajectory-grouped cross-validation, deep-ensemble predictions, and Monte Carlo dropout are used to assess model robustness and predictive uncertainty. Sensitivity analyses demonstrate that physics constraints can be incorporated over a broad range of loss weights without compromising predictive accuracy, while substantially improving compliance with the prescribed physical trends. Benchmarking against conventional machine-learning regressors further demonstrates competitive predictive performance. SHAP analysis across the cross-validation ensemble identifies the coupled near-icosahedral motif family, particularly the full icosahedral motif and its single-atom-perturbed counterpart, as the dominant structural fingerprints of the glassy state. These results demonstrate that physically motivated constraints can be embedded directly into motif-based structure–property models, providing a pathway from post-hoc interpretability toward physically constrained machine learning for metallic glasses.

## 1. Introduction

Metallic glasses (MGs) occupy a distinctive position among structural and functional materials: they combine the high strength, large elastic limits, and corrosion resistance associated with disordered atomic packing with the processability of a viscous liquid [1]. Unlike their crystalline counterparts, MGs possess no long-range translational order, and consequently their macroscopic thermal, mechanical, and transport properties are dictated entirely by the statistics of local atomic environments rather than by a well-defined unit cell [2,3]. Establishing a quantitative link between these local structural motifs and the thermodynamic, kinetic, and mechanical state of a glass has remained a central unresolved problem in metallic-glass science for more than two decades [4]. Part of the difficulty is that no single structural descriptor — pair correlation functions, radial distribution functions, Voronoi polyhedron statistics, flexibility volume, configurational entropy, or boson-peak intensity fully captures the relevant physics on its own; each illuminates a different facet of a fundamentally multidimensional structure–property landscape [3,5–7]. Among the binary glass-forming systems studied computationally and experimentally, Cu–Zr has emerged as the paradigmatic model alloy for interrogating this structure–state relationship [2,8,9]. A substantial body of work has established that Cu-centered full icosahedra (FI), identified by the Voronoi index ⟨0,0,12,0,0⟩ with coordination number 12, dominate the short-range order of Cu-rich Cu–Zr glasses. These clusters grow in population during structural relaxation and physical aging, percolate into a mechanically stiff interpenetrating backbone as the supercooled liquid approaches its glass-forming composition range, and act to suppress local atomic mobility[8–14]. Advances in atomic electron tomography and fluctuation electron microscopy have since provided direct experimental confirmation of individual icosahedral clusters and their higher-order superclusters in both Zr-rich and multicomponent metallic glasses [15,16]. Full icosahedra are furthermore understood as the structural antithesis of "soft spots" — geometrically unfavored motifs (GUMs) that preferentially host the quasi-localised low-frequency vibrational modes from which shear transformation zones nucleate under mechanical load [17,18]. This body of evidence converges on a clear physical prior: the FI is the canonical local motif of Cu–Zr glass formation, and it is the percolating icosahedral network, not the isolated FI population in isolation, that constitutes the most physically defensible structural proxy for the glassy state. A complementary, data-driven perspective on the Cu–Zr motif vocabulary has emerged more recently. [19] catalogued the full spectrum of local Cu-centered environments across compositions and temperatures. They demonstrated that approximately 90% of them can be grouped into a compact set of icosahedral and near-icosahedral "building blocks," alongside a small number of non-icosahedral outlier motifs whose population grows systematically as the glass is deepened kinetically [19]. In a related direction, [20] applied a machine-learning search over the full Voronoi-motif space of Cu–Zr molecular-dynamics configurations and uncovered a previously unrecognized "vacancy-like" structural motif, denoted Q7 (a Voronoi polyhedron with seven quadrangular faces), whose population follows an Arrhenius-like temperature dependence reminiscent of classical vacancy thermodynamics in crystalline metals and correlates strongly with yielding and local failure during mechanical deformation. Q7 represents the disordered extremum of the Cu–Zr motif spectrum: a locally under-packed environment in which a notional structural "hole" can be statistically defined, and its discovery demonstrates that supervised machine learning applied directly to atomistic Voronoi statistics can recover physically meaningful local environments that decades of hand-crafted descriptor design had not isolated. This motif-discovery paradigm has since been extended architecturally, [21] developed a graph convolutional neural network (GCNN) classifier operating directly on atomistic graphs to identify glassy states in metallic glasses, offering a complementary route to motif-population regression that instead learns the decision boundary between structural classes directly from atomic connectivity. The rise of machine learning across computational materials science more broadly has paralleled and substantially accelerated these motif-level advances. Machine-learning interatomic potentials (MLIPs) now permit near-DFT-accuracy simulations of glass-forming systems at scales inaccessible to ab initio methods. For the Cu–Zr system specifically, [22] developed a general-purpose MLIP based on the atomic cluster expansion (ACE) formalism, trained against an extensive density-functional-theory dataset spanning both crystalline and amorphous phases, and demonstrated that it reproduces the experimental phase diagram and short-range icosahedral order with markedly improved fidelity relative to classical embedded-atom potentials. Comparative benchmarking of MLIP families accuracy, computational speed, and data efficiency has further clarified the practical trade-offs facing modelers of disordered metallic systems [9]. In parallel, supervised learning has been used extensively to predict glass-forming ability (GFA) and critical cooling rates from compositional and simulated structural features; [23] showed that a 201-feature machine-learning model for Cu–Zr–Al GFA prediction was statistically indistinguishable from a model trained on deliberately unphysical features and was in fact outperformed by a three-feature, human-designed baseline a cautionary result that motivated the explicit call for physically grounded, explainable machine-learning pipelines in metallic-glass informatics. Building on this, [24] combined gradient-boosted learners with SHapley Additive exPlanations (SHAP) at the alloy-composition level to attribute GFA predictions to interpretable compositional descriptors, while related critical-cooling-rate models have used SHAP to confirm that the most influential simulated features, energies above the convex hull, heat-capacity changes, and icosahedra-like Voronoi fractions, carry physically reasonable directions of influence. SHAP itself, introduced by [25] as a unifying, game-theoretically grounded framework for additive feature attribution has become the default tool for extracting post-hoc interpretability from otherwise opaque deep models in materials informatics, including its first application directly to individual-configuration motif-level features in Cu–Zr glasses, which established that a tightly coupled near-icosahedral family (coordination numbers 11–13, anchored by the FI and its single-atom-perturbation sibling ⟨1,0,9,3,0⟩) collectively encodes the thermodynamic state of the system across the full liquid–supercooled–glass range. A structurally distinct but physically related research direction concerns nanoglasses (NGs) glasses synthesized via a bottom-up consolidation of nanometer-scale amorphous clusters rather than by conventional melt-quenching. [26] articulated the nanoglass concept as a new class of noncrystalline material comprising glassy grains separated by glass–glass interfaces of locally enhanced free volume and modified short-range order. [27] characterized the atomic structure, thermal stability, and nanoindentation response of consolidated Cu50Zr50 nanoglasses, revealing pronounced Cu/Zr elemental segregation between glassy cores and Cu-enriched interfacial regions that governs both crystallization kinetics and mechanical performance. Molecular-dynamics studies of amorphous–crystalline nanolaminates and nanoglass composites have further shown that martensitic phase transformations and interfacial free volume jointly control shear-band nucleation and stabilization in these heterogeneous glassy architectures [28]. [29] contributed early atomistic evidence in this space as well, using large-scale molecular dynamics to examine mechanical failure and structural heterogeneity in nanostructured amorphous metallic systems. Collectively, this nanoglass literature reinforces the same physical conclusion emerging from the bulk Cu–Zr motif-discovery work: interfacial [30] and boundary-adjacent local environments, whether they are the geometrically unfavored GUMs that host soft spots in bulk glasses or the free-volume-rich glass–glass interfaces of nanoglasses, are the loci where structural disorder couples most directly to mechanical response.

Despite this convergence of evidence from motif statistics, explainable machine learning, and nanoglass interface science, a persistent limitation runs through nearly all of the data-driven structure–property work reviewed above: the learned models are trained purely from data, with physical knowledge entering, if at all, only as a post-hoc explanatory lens (via SHAP or permutation importance) rather than as a constraint shaping the learned function itself. This leaves open the possibility that a model's apparent physical consistency is an artifact of the specific training distribution rather than a property enforced by design, and it offers no guarantee that the model will extrapolate sensibly beyond the sampled composition–quench-rate space. Physics-informed neural networks (PINNs), introduced by [31] as a framework for embedding governing differential-equation residuals directly into the training loss of a neural network, provide a natural avenue for closing this gap: rather than discovering physical consistency after the fact, one can impose it as a soft constraint during optimization, using autograd-computed sensitivities of the network's own output with respect to its inputs. This approach has been applied extensively in fluid mechanics and solid mechanics but has seen comparatively little application to the motif-population regression problem at the heart of metallic-glass structure–state characterization.

In this work, we address these gaps via expanding on the recently developed Voronoi-motif regression and SHAP-attribution framework for Cu–Zr metallic glasses, we create a physics-informed neural network that applies autograd-based gradient-residual penalties during training to enforce two physically relevant constraints: (i) a monotonicity restriction on the near-icosahedral motif family aligned with its recognized role as a stable-glass structural indicator, and (ii) a causal-sign constraint that connects the network's temperature prediction to the logarithm of the quench rate, capturing the physically anticipated fictive-temperature signature of kinetically distinct cooling trajectories. We also broaden the framework to a multi-task context where an additional classification head simultaneously forecasts the thermodynamic state (liquid, supercooled/transition, or glass) from the identical motif-population feature vector. Additionally, we assess predictive uncertainty through both deep-ensemble and Monte Carlo-dropout estimators, which are evaluated using trajectory-disjoint five-fold cross-validation to avoid information leakage between molecular-dynamics trajectories. Together, these methodological additions aim to convert a previously purely explanatory pipeline into one whose physical consistency is a property of the trained model itself, a distinction we argue is essential as motif-informed machine learning moves from post-hoc structural interpretation toward predictive and, ultimately, inverse-design applications in metallic-glass science.

## 2. Methodology

### 2.1. Dataset preparation

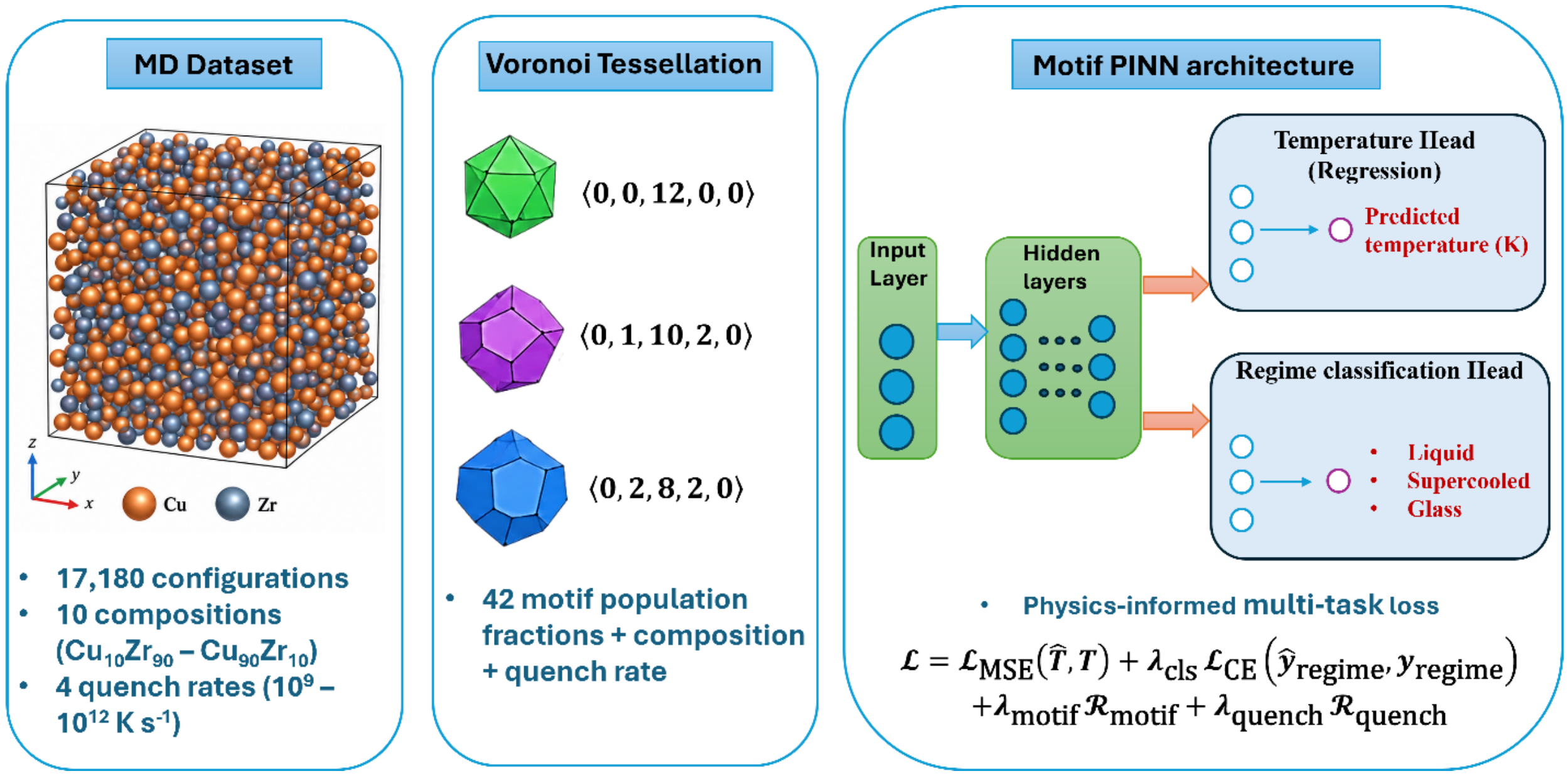

***Figure 1.*** *Schematic workflow of the proposed physics-informed multi-task learning framework. Molecular-dynamics simulations generate 17,180 atomic configurations spanning 10 Cu–Zr compositions and four quench rates ($10^9$–$10^{12}$ K s$^{-1}$). Voronoi tessellation is subsequently used to quantify local atomic environments through 42 motif population fractions, which are combined with composition and quench rate as model inputs. The resulting descriptors are supplied to a shared neural-network architecture with two task-specific heads for temperature regression and thermal-regime classification.*

To address the coupled influence of local atomic structure, composition, and thermal history on the thermal behaviour of Cu–Zr metallic glasses, a physics-informed multi-task learning framework was developed, as schematically illustrated in Figure 1. All molecular-dynamics-derived structural data used in this study were obtained from the publicly archived dataset [32]. The dataset comprises 17,180 atomistic configurations spanning 10 Cu–Zr compositions ($Cu_{10}Zr_{90}$–$Cu_{90}Zr_{10}$), generated using molecular dynamics with an embedded-atom-method potential by quenching the systems from 2000 to 50 K at four constant quench rates ranging from $10^9$ to $10^{12}$ K $s^{-1}$. For every configuration, three-dimensional Voronoi tessellation was applied to define the local atomic environments based on the population fractions of the 100 most prevalent Voronoi motifs, along with the Cu concentration, the base-10 logarithm of the quench rate, the identifier of the source MD trajectory, and the instantaneous temperature. In this study, the archived dataset was utilized as-is as the ground truth for training, validating, and testing the model, with no extra molecular-dynamics simulations conducted. The 42 most common Voronoi motif population fractions were selected from the available structural descriptors as structural features and combined with composition and quench rate to form the model input space. These descriptors were input into a common neural network representation featuring two output heads tailored for specific tasks: one for temperature regression to continuously predict temperature and another for classifying thermal regimes, categorizing each configuration as liquid, supercooled-transition, or glass. Both tasks were trained at the same time utilizing a physics-informed multi-task objective that merges regression and classification losses along with regularization terms linked to the structural motifs and quench rate. This framework thus combines atomic structural details with thermal-processing history while concurrently learning the continuous temperature reaction and the associated thermodynamic regime.

**2.2. Feature construction**

Raw per-configuration motif counts were converted to fractional populations by normalizing each of the 100 tabulated motif counts by their row-wise sum. Following the reference study, the 40 most prevalent motifs (by mean fractional population across the full dataset) were retained as regression features; any of the six physically constrained near-icosahedral motifs discussed in Section 2.4 that fell outside this top-40 set were force-included to ensure the physics-informed loss terms (Section 2.4) had access to all relevant inputs. The Cu concentration and the $\log10_{10}$-quench-rate were appended as two additional scalar features, yielding a feature vector of dimensionality consistent with the 42-input representation used in the reference study. Each configuration was additionally assigned a categorical thermal-regime label — glass ($T < 300$), supercooled/transition ($300 \leq T < 1500$), or liquid ($T \geq 1500$) — derived from its temperature label, using boundaries matching the qualitative regime discussion in the reference paper, for use as an auxiliary classification target (Section 2.3).

### 2.3. Network architecture

A multi-task, physics-informed feed-forward neural network (PINN) was implemented in PyTorch. A shared trunk of three hidden layers (32, 64, and 16 rectified-linear units at the default width multiplier, matching the architecture of the reference study) feeds two independent linear output heads: (i) a regression head producing a single scalar, the predicted temperature $\hat{T}$, and (ii) a three-way classification head producing logits over the glass/supercooled-transition/liquid regime labels. Both heads share the same learned representation, so the network is trained jointly rather than as two separate models. Dropout was optionally applied after each hidden layer, disabled (probability 0) for the deep-ensemble models and enabled at $p = 0.10$ for the Monte Carlo (MC)-dropout uncertainty-quantification models.

### 2.4. Physics-informed loss terms

In addition to the standard supervised losses, mean-squared error for the regression head and cross-entropy for the classification head, two physics-residual penalty terms were computed via automatic differentiation of the network's own regression output with respect to its inputs, and added to the total training loss:

$$\mathcal{L} = \mathcal{L}_{\text{MSE}}\left(\hat{T}, T\right) + \lambda_{\text{cls}}\, \mathcal{L}_{\text{CE}}\left(\hat{y}_{\text{regime}}, y_{\text{regime}}\right) + \lambda_{\text{motif}}\, \mathcal{R}_{\text{motif}} + \lambda_{\text{quench}}\, \mathcal{R}_{\text{quench}}$$

**(A) Near-icosahedral monotonicity constraint.**

For the full icosahedron $\langle 0,0,12,0,0\rangle$ and four of its coordination-number 11–13 near-icosahedral siblings ($\langle 0,2,8,1,0\rangle$, $\langle 0,2,8,2,0\rangle$, $\langle 0,1,10,2,0\rangle$, $\langle 0,2,8,5,0\rangle$, the gradient residual $\mathcal{R}_{\text{motif}}$ penalizes any positive sensitivity $\partial\hat{T}/\partial f_i > 0$ enforcing the physically expected stable-glass signature (rising near-icosahedral population should not raise the predicted temperature).

For the single-atom-perturbation motif $\langle 1,0,9,3,0\rangle$, identified in the reference study as a high-temperature precursor/soft-spot signature, the sign of the penalty is reversed, penalizing $\partial\hat{T}/\partial f_i < 0$. Each violation is penalized as a squared-ReLU residual, $[\text{ReLU}(\pm g_i)]^2$ averaged over the training batch.

**(B) Quench-rate causality constraint.**

A second gradient-residual term penalizes any negative sensitivity of the predicted temperature to log-quench-rate, $\partial\hat{T}/\partial(\log_{10}\dot{q}) < 0$ reflecting the physical expectation, motivated directly by the reference paper's discussion of the residual "quench-rate signal" needed to disambiguate structurally degenerate configurations, that faster-quenched configurations are frozen at a higher fictive temperature at fixed motif population.

Both residuals are computed such that their gradients with respect to the network parameters can themselves be backpropagated (a genuine PINN gradient-penalty term, not a post-hoc explanation), and gradients are clipped to a maximum norm of 5.0 during optimization for numerical stability.

### 2.5. Training protocol, data splitting, and hyperparameter selection

**Trajectory-disjoint holdout.** To prevent information leakage between configurations sampled from the same MD trajectory, the 17,180 configurations were first partitioned into 80% development and 20% test groups (approximately 20% of unique trajectories held out), using a fixed random seed; the test set was untouched until final evaluation.

**Cross-validation.** The 80% development pool was evaluated with 5-fold cross-validation (grouped by trajectory), following the same protocol as the reference study, so that no MD trajectory is split across the training and validation portions of a fold. Feature and target scalers were fit exclusively on each fold's training partition and applied to that fold's validation partition, avoiding cross-fold leakage.

**Hyperparameter fine-tuning.** A random search (30 sampled configurations) was performed over network width multiplier, learning rate, the two physics-penalty weights ($\lambda_{\text{motif}}$, $\lambda_{\text{quench}}$), the auxiliary classification weight ($\lambda_{\text{cls}}$), and dropout probability, using an inner 3-fold split of the development pool only; the 20% trajectory-disjoint test set was never used for hyperparameter selection. Distinct random seeds were used to control the trajectory holdout, the hyperparameter search, and model-weight initialization independently, so that changes to the tuning grid do not perturb the composition of the outer test split.

**Physics-weight sensitivity analysis.** Following hyperparameter selection, one-at-a-time sweeps of $\lambda_{\text{motif}}$ and $\lambda_{\text{quench}}$ (over the range 0–0.5), together with an explicit four-way ablation (no physics / motif-only / quench-only / both constraints), were each evaluated by full 5-fold cross-validation on the development pool, reporting mean $\pm$ standard deviation $R^2$, RMSE, and MAE across folds.

**Optimization.** Networks were trained with the Adam optimizer for up to 700 epochs, batch size 128, with early stopping (patience of 50 epochs, monitored on regression validation MSE) as the primary stopping criterion; classification validation loss was tracked in parallel as a secondary diagnostic but did not govern checkpoint selection.

## 2.6. Auxiliary regime classification, feature attribution, and classical benchmarks

The classification head's test-set predictions (ensemble-averaged class probabilities) were evaluated against the true thermal-regime labels via a confusion matrix and standard classification metrics (precision, recall, F1). Global feature importance for the regression task was assessed using SHAP GradientExplainer, averaged across all five cross-validation models and across multiple independently sampled training backgrounds ($n = 3$ backgrounds of 150 samples each) for stability; where SHAP was unavailable, permutation importance was substituted. Finally, the PINN was benchmarked against (i) an architecture-matched baseline neural network trained with both physics weights set to zero, and (ii) three classical regressors, random forest, degree-2 polynomial ridge regression, and (where available) gradient-boosted trees (XGBoost/CatBoost) all evaluated under the identical trajectory-grouped 5-fold cross-validation protocol on the development pool and on the same untouched 20% test set, to isolate the contribution of the physics-informed loss terms from that of model architecture alone.

## 3. Results & Discussions

### 3.1 PINN training and convergence

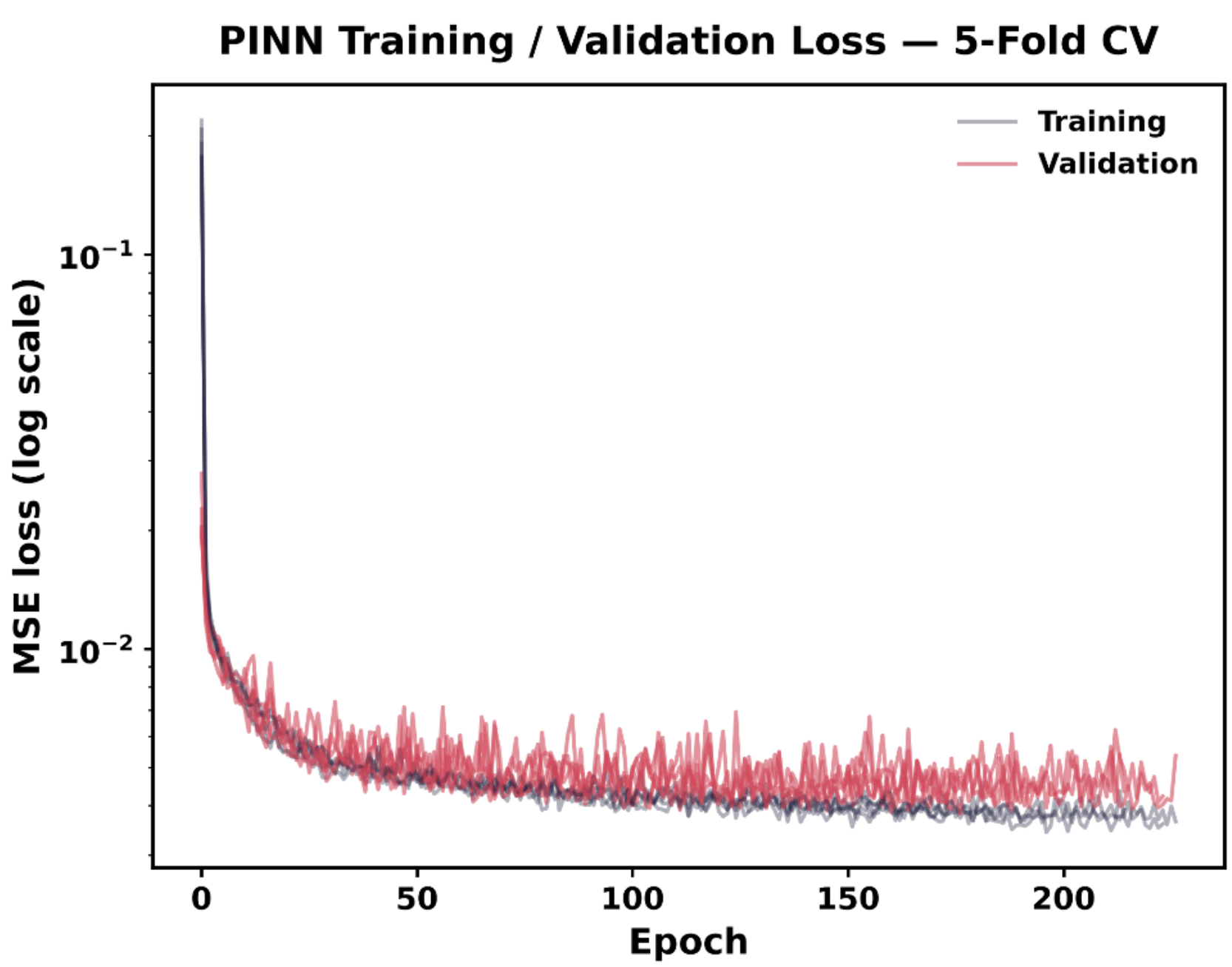

***Figure 2.*** *Training and validation mean-squared-error loss of the multi-task, multi-physics PINN across five GroupKFold cross-validation folds (grouped by MD trajectory). Both losses decrease rapidly and plateau without a growing train-validation gap, indicating stable optimization and no over-fitting.*

The convergence characteristics of the suggested multi-task PINN were assessed via five-fold trajectory-grouped cross-validation, while the training and validation mean-squared-error (MSE) losses were tracked continuously during the optimization process. Figure 2 displays the related loss histories on a logarithmic scale for all five folds. The model demonstrates a swift decline in loss during the early training phase, subsequently showing a gradual reduction as the optimization nears a stable state. In the initial several tens of epochs, the losses drop by over an order of magnitude, and subsequently, both training and validation losses stay within the low-$10^{-3}$ range. Significantly, the validation losses align with the general pattern of the training losses without any prolonged divergence in the later epochs. Even though the validation curves show more variations than the training curves, their levels stay fairly consistent in the later phases of training. This behavior shows that the network can generalize across the validation subsets that are separated by trajectories instead of merely recalling specific molecular-dynamics trajectories. The minimal gap between the training and validation losses indicates that the integration of regression, classification, and physics-informed objectives does not lead to significant instability in optimization. The swift initial decline in loss is succeeded by a plateau phase, suggesting that the key predictive insights from the Voronoi-motif population descriptors are acquired fairly soon during training. Later epochs mainly offer gradual enhancement of the model parameters and the related physics-constrained components. The uniform convergence behavior observed in all five folds further reinforces the reliability of the training process and establishes a foundation for the later assessment of predictive accuracy, uncertainty, and physics alignment.

**3.2 Predictive performance of the motif-PINN**

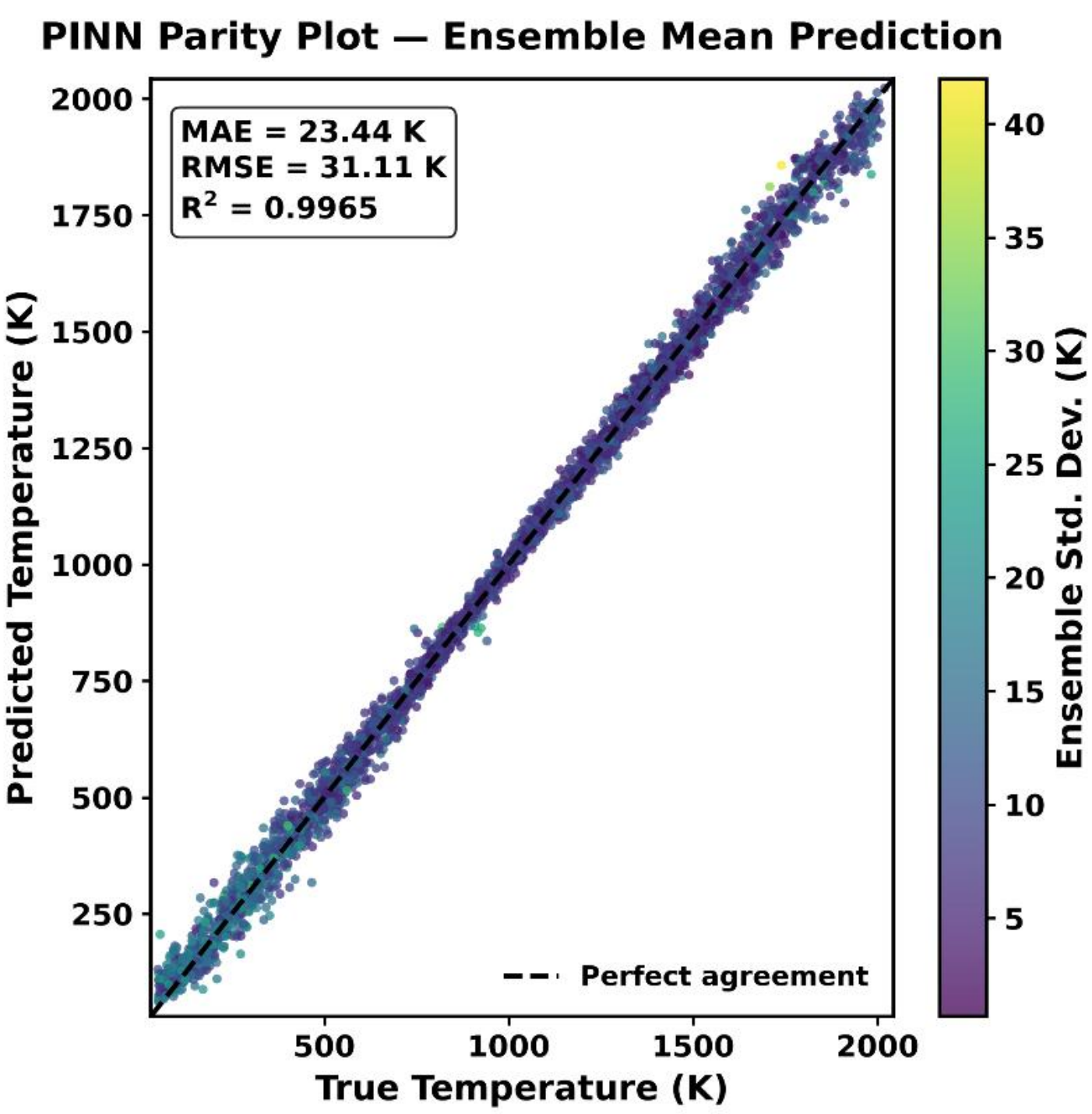

***Figure 3.*** *Parity plot of predicted versus true temperature on the held-out, trajectory-disjoint test set, using the deep-ensemble mean prediction from the five cross-validation folds. Point color encodes the ensemble standard deviation (epistemic uncertainty). MAE = 23.44 K, RMSE = 31.11 K, R2 = 0.9965.*

The predictive capability of the proposed motif-based PINN was first evaluated using ensemble-mean predictions on the held-out data. Figure 3 presents the parity plot between the predicted and true configuration temperatures, with the dashed diagonal representing perfect agreement. The predictions closely follow the $T_{pred} = T_{true}$ line over the full temperature range considered, extending from approximately 50 to 2000 K. This close correspondence demonstrates that the Voronoi-motif population histograms contain sufficient structural information to accurately infer the thermodynamic state of the Cu–Zr metallic-glass configurations.

The ensemble-mean prediction achieves an MAE of 23.44 K, RMSE of 31.11 K, and $R^2 = 0.9965$. The modest difference between MAE and RMSE suggests that the total error isn't significantly influenced by a few large outliers. Additionally, the elevated $R^2$ value indicates that the model accounts for almost all the variation in configuration temperature throughout the dataset. A solid consensus is upheld throughout the low- and high-temperature segments of the dataset, with no significant systematic bias apparent in the parity distribution. The color scale gives extra details about the predictive uncertainty derived from the ensemble. Every point is colored based on the standard deviation of the predictions from the individual trained ensemble members. Most configurations display relatively minor ensemble variations, indicating uniform predictions among the ensemble members. Higher uncertainty values arise for a smaller selection of configurations and are typically linked to points showing a more significant deviation from the perfect-agreement line. This offers an initial insight that ensemble disagreement holds valuable information about prediction reliability, which will be quantitatively analyzed later through the uncertainty–error assessment. In summary, these findings demonstrate that the suggested PINN delivers very precise temperature predictions based on local structural details while ensuring prediction consistency throughout the wide thermodynamic spectrum reflected in the molecular-dynamics dataset.

### 3.3 Residual and statistical validation

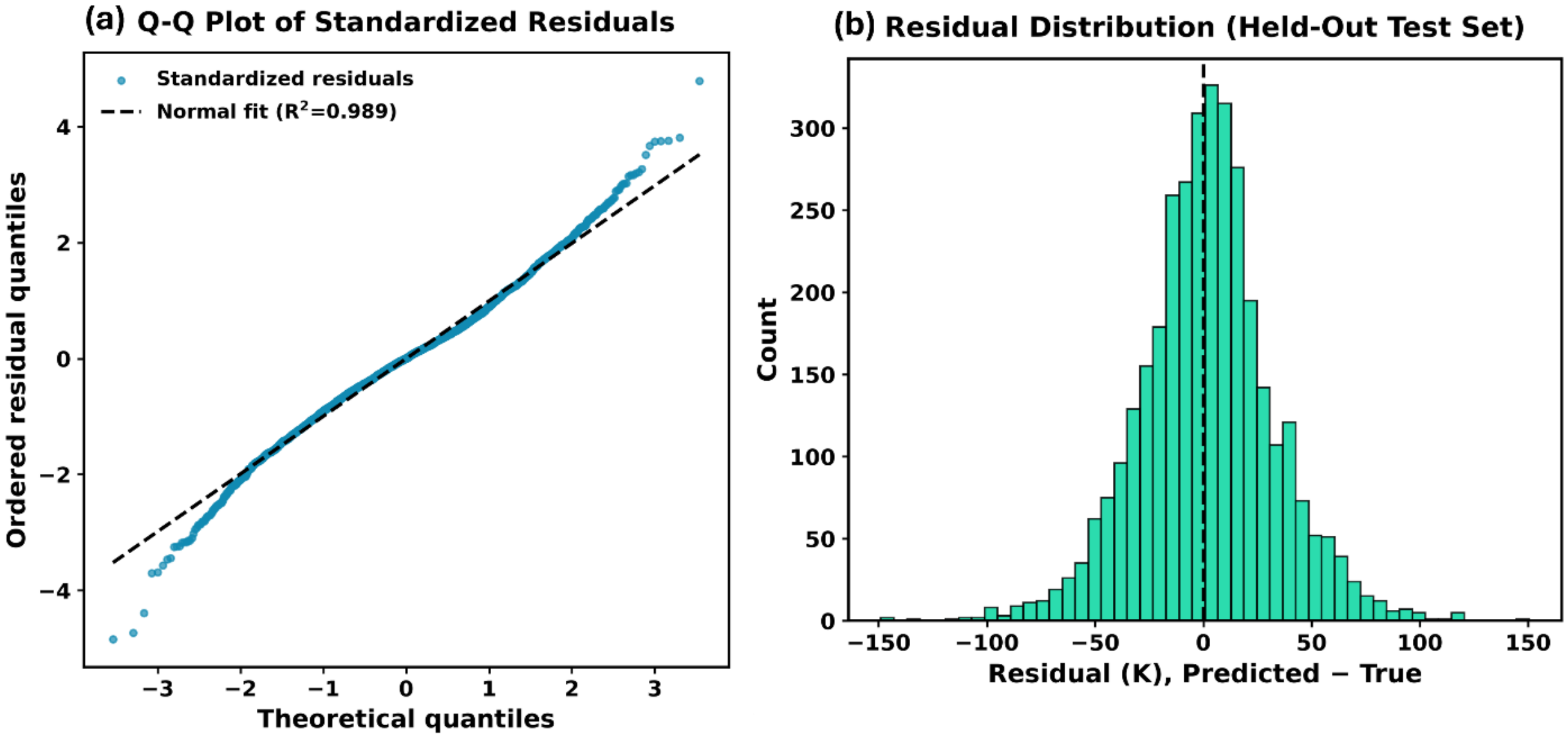

***Figure 4.*** *(a) Quantile-quantile plot of standardized test-set residuals against the standard normal distribution. The central quantiles show approximately Gaussian behavior, while deviations in the tails indicate departures from strict normality. (b) Distribution of test-set residuals (predicted minus true temperature). The approximately symmetric, zero-centered distribution indicates no systematic bias in the PINN predictions across the operating temperature range.*

The statistical features of the prediction errors were additionally analyzed through a Q–Q plot of the standardized residuals (Fig. 4a). The residuals closely adhere to the fitted normal trend in the central part of the distribution, aligning with an approximately Gaussian error structure. Nonetheless, significant variations arise in the tails of the distribution, featuring multiple instances where standardized residual values are near five. Consequently, although most of the residual distribution resembles a normal distribution, the presence of heavier tails suggests that a purely Gaussian assumption could underestimate the likelihood of somewhat large prediction errors. This observation is taken into account when understanding the uncertainty estimates provided later.

The residual distribution provides further evidence of the high predictive accuracy of the PINN (Fig. 4b). The residuals, defined as $T_{\text{pred}} - T_{\text{true}}$, are tightly grouped around zero, with most errors falling within roughly ±50 K. The proximity to the zero center suggests that the model does not show a significant systematic bias on the held-out dataset. Still, the distribution shows unequal and elongated tails, with a few predictions straying by over 100 K. This pattern aligns with the deviations from the normal reference line noted in the Q–Q analysis and suggests that the residual distribution is roughly Gaussian instead of being perfectly normal.

### 3.4 Thermal-Regime Classification Performance

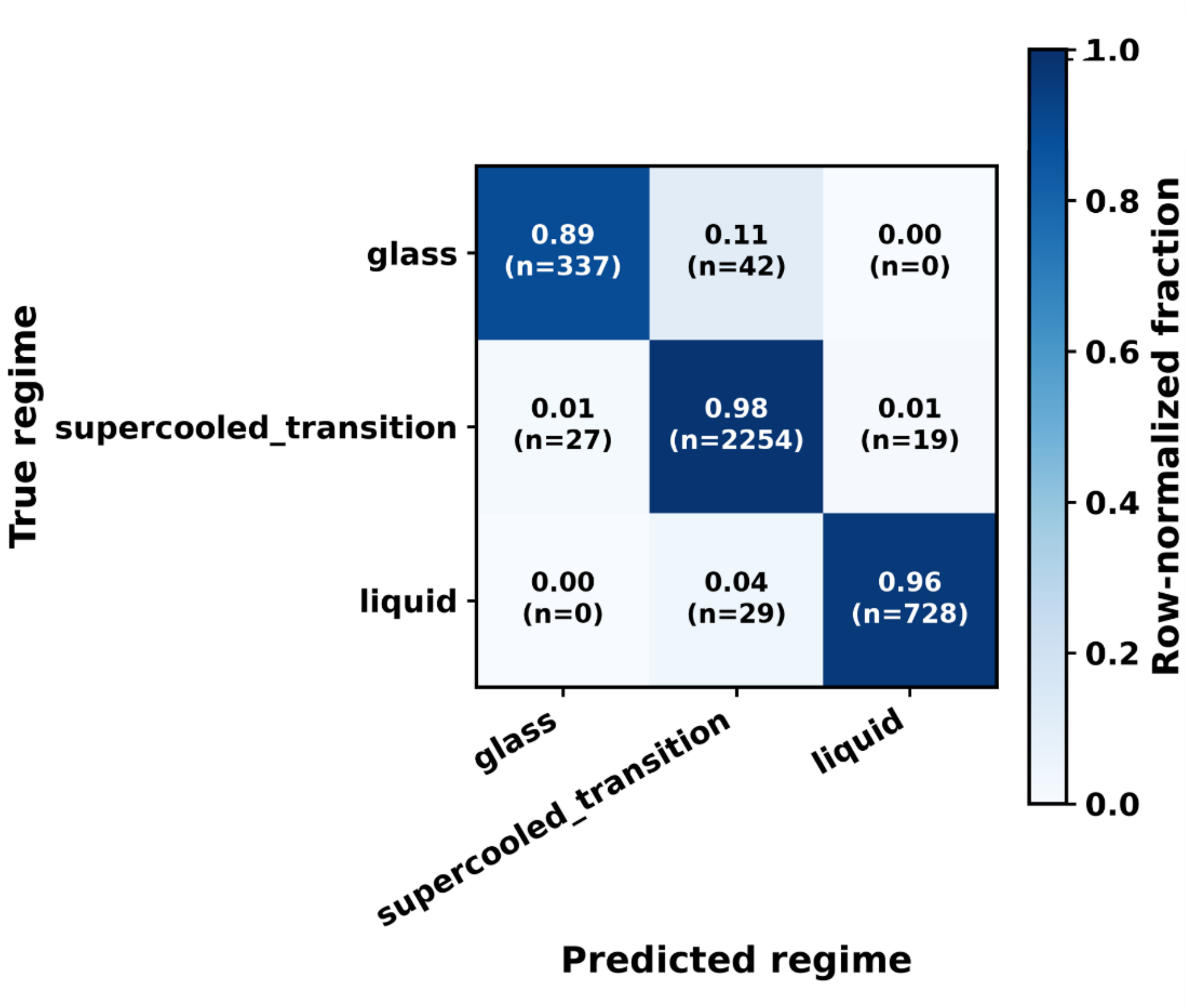

***Figure 5.*** *Confusion matrix for the auxiliary thermal-regime classification head (glass: T<300 K; supercooled/transition: 300-1500 K; liquid: T>1500 K), evaluated on the held-out test set using the ensemble-averaged class probabilities. Diagonal dominance indicates the same motif-population features that support accurate temperature regression also linearly separate the three thermodynamic regimes.*

The performance of the auxiliary thermal-regime classification task is evaluated using the row normalized confusion matrix shown in Fig. 5. The model exhibits strong classification capability across all three regimes, with an overall accuracy of 96.6% and a macro-F1 score of approximately 0.95. The model classifies the supercooled-transition regime most accurately, with 98% of configurations correctly assigned to this class, while the corresponding correct-classification fractions for the glass and liquid regimes are 89% and 96%, respectively. Importantly, the misclassifications are concentrated almost exclusively between adjacent thermal regimes. Among the glass configurations, 11% are assigned to the supercooled-transition regime, whereas 4% of liquid configurations are similarly classified as supercooled transition. In contrast, no direct confusion is observed between the glass and liquid regimes. This distribution of classification errors is physically consistent with the continuous nature of the underlying thermal-state evolution, where the distinction between glass and supercooled transition, or between supercooled transition and liquid, is expected to be less distinct than that between the two limiting states. The confusion matrix therefore indicates that the classification head captures meaningful thermal-regime boundaries rather than producing indiscriminate classification errors.

### 3.5 Voronoi-motif populations

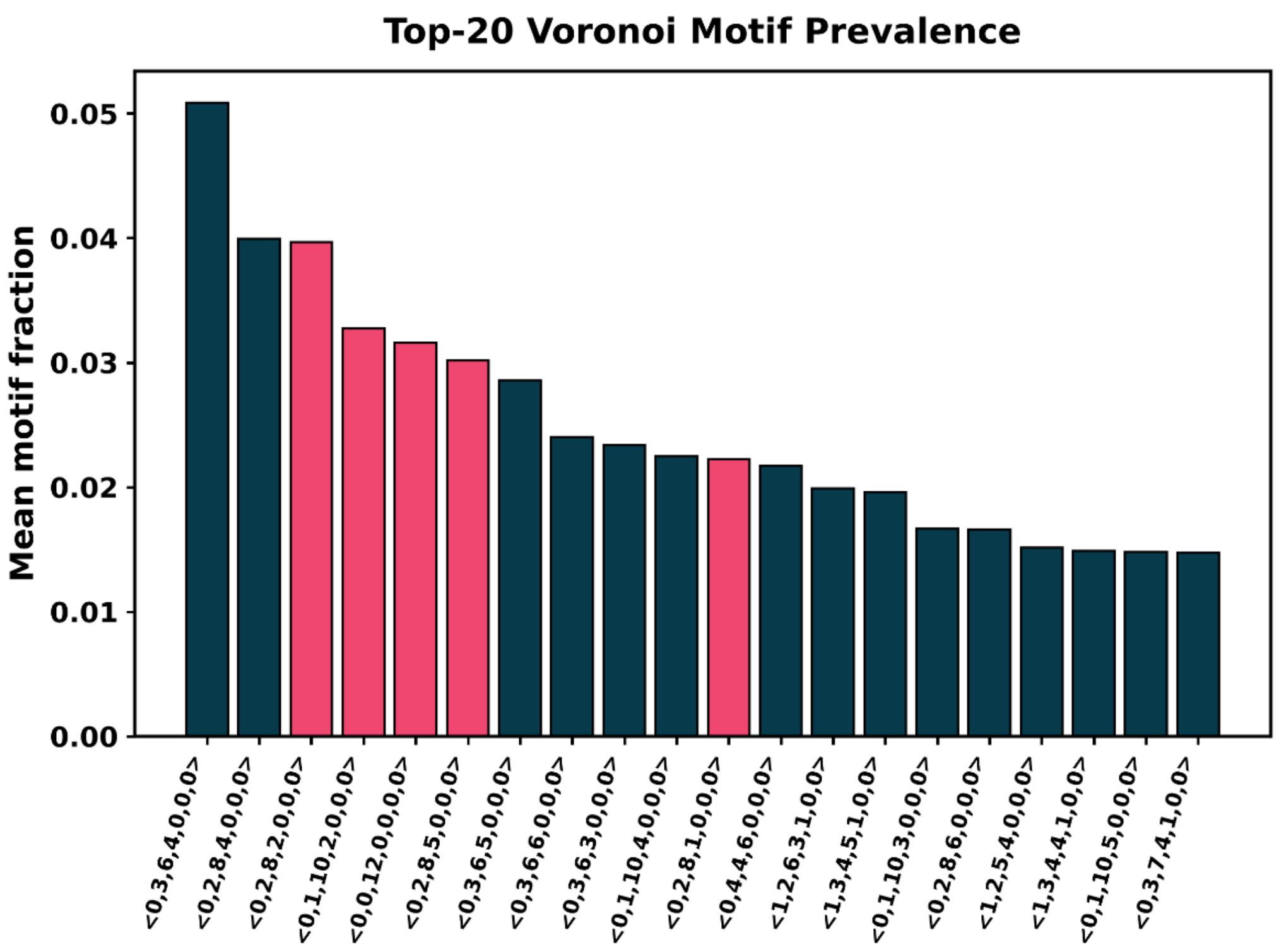

***Figure 6.*** *Mean fractional population of the twenty most prevalent Voronoi motifs across the full dataset. Motifs highlighted in red are members of the near-icosahedral family subject to the motif-monotonicity physics constraint.*

Figure 6 presents the mean population fractions of the 20 most prevalent Voronoi motifs across the Cu–Zr metallic-glass configurations. The population of the individual motifs ranges from approximately 1.5% to 5.1%, indicating that the local structure is distributed among a relatively broad spectrum of polyhedral environments rather than being dominated by a single configuration. The most frequently occurring motif accounts for approximately 5.1% of the atomic environments, while several other motifs exhibit populations between 2 and 4%. Notably, multiple members of the near-icosahedral structural family, highlighted in pink, occur among the dominant motifs. Their substantial population is consistent with the prevalence of five-fold-related local ordering in Cu–Zr metallic glasses. However, motif population alone does not establish its relevance to the thermodynamic state. Consequently, the motif frequencies shown here are subsequently compared with model-derived SHAP attributions to distinguish structural abundance from predictive importance. This comparison provides the basis for identifying whether the motifs that are most common in the glass are also those that carry the strongest information regarding configuration temperature.

### 3.6 Structural interpretation of the PINN using ensemble SHAP analysis

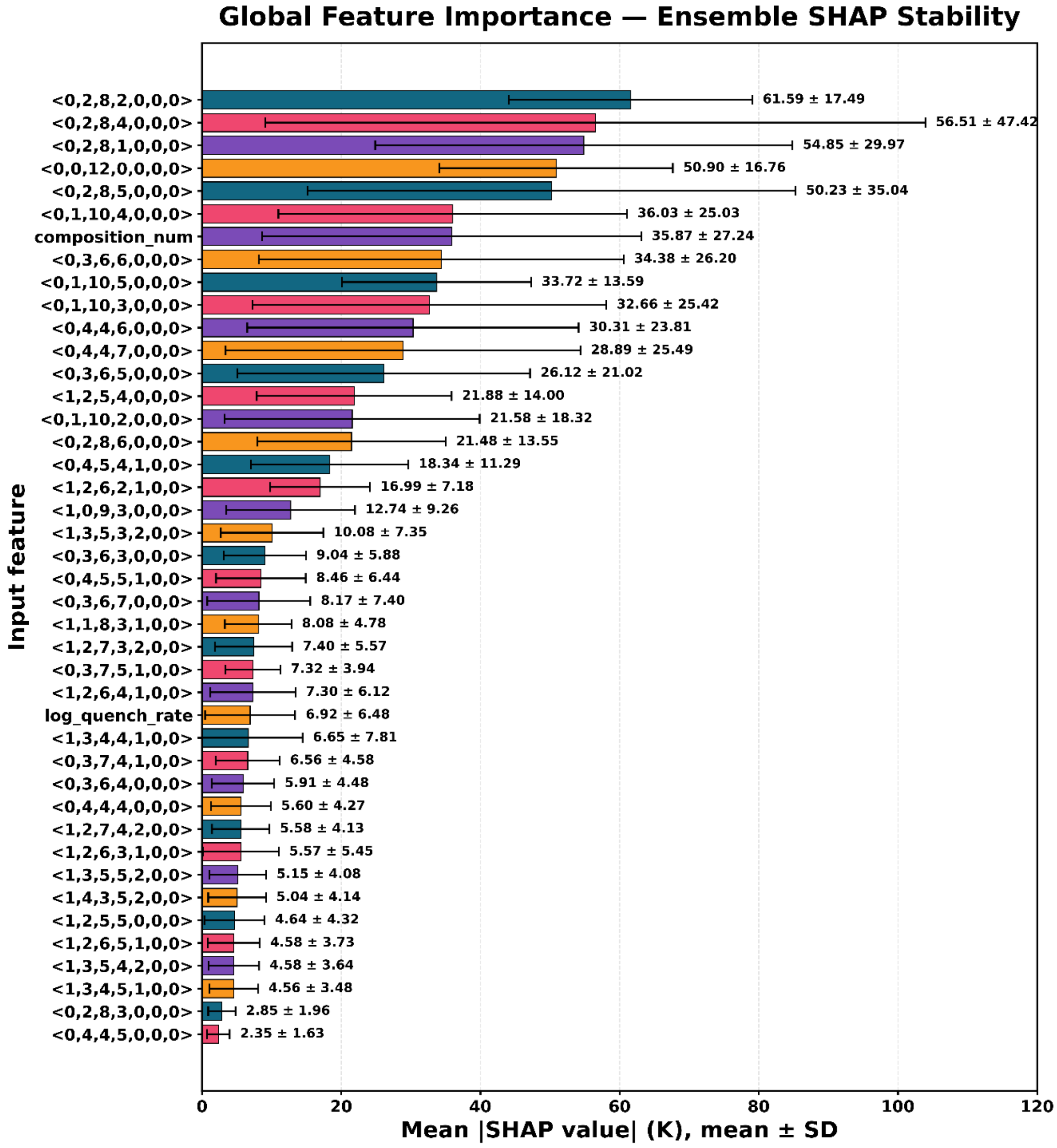

***Figure 7.*** *Global feature importance for the regression head based on mean absolute SHAP values from GradientExplainer. To assess robustness, SHAP values were evaluated for 5 cross-validation models using 3 independently sampled training-fold backgrounds per model; bars show mean |SHAP| and error bars show the standard deviation across model/background runs. The final test set was used only as the explanation target and never as a SHAP background.*

Figure 7 presents the global feature importance obtained from the ensemble SHAP analysis of the PINN temperature-regression task. Feature importance is quantified as the mean absolute SHAP value, expressed in kelvin, with the error bars representing the standard deviation of the absolute SHAP values across the ensemble predictions. The ranking demonstrates that the predicted temperature is governed predominantly by local atomic structural motifs, with the most influential descriptors being the Voronoi motifs $\langle 0, 2, 8, 2, 0, 0, 0 \rangle$, $\langle 0, 2, 8, 4, 0, 0, 0 \rangle$, and $\langle 0, 2, 8, 1, 0, 0, 0 \rangle$, which exhibit mean absolute SHAP values of 61.59, 56.51, and 54.85 K, respectively. The perfect icosahedral motif $\langle 0, 0, 12, 0, 0, 0, 0 \rangle$ demonstrates a significant overall contribution of 50.90 ± 16.76 K, emphasizing the crucial role of well-coordinated local settings in influencing the expected thermal condition.Composition ranks as one of the key predictors, with an average absolute SHAP value of 35.87 ± 27.24 K, suggesting that chemical composition plays a significant role in temperature prediction alongside local structural details. Conversely, the logarithm of the quench rate shows a significantly lower global contribution of 6.92 ± 6.48 K. This suggests that, in the current dataset, the immediate local structural condition and composition offer more reliable direct insights for temperature forecasting than the nominal processing history indicated by the quench rate. Significantly, the considerable standard deviations noted for various prominent motifs suggest that their impacts are inconsistent across different configurations, fluctuating with composition and thermal condition. Consequently, the ensemble SHAP analysis reveals that local Voronoi structure is the main source of predictive information, while also showing the interdependent effects of composition and thermal history.

### 3.7 Interpretability of motif-based temperature prediction

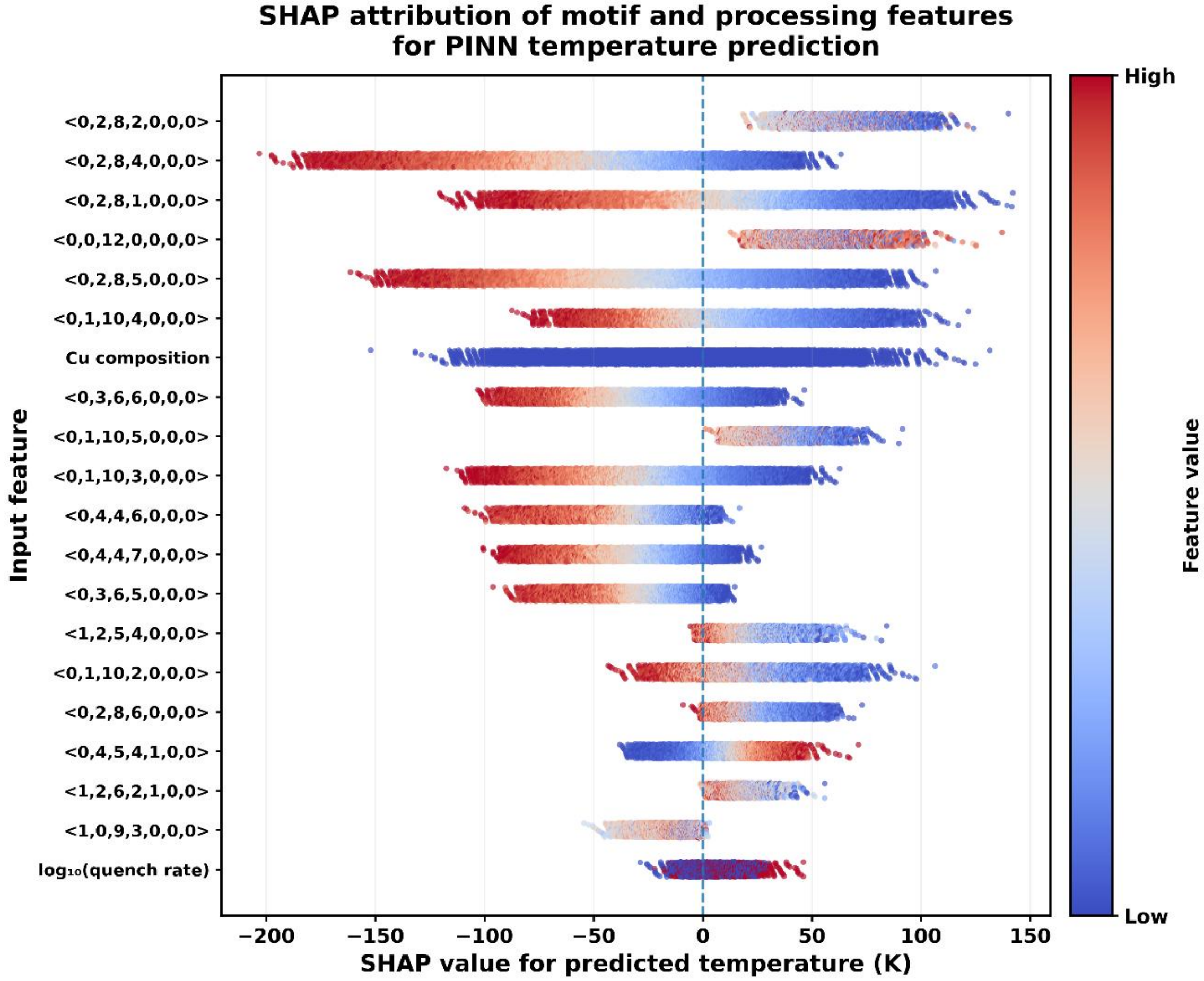

*Figure 8. SHAP beeswarm plot showing the contribution of Voronoi-motif populations, Cu composition, and quench rate to the predicted temperature. Each point represents one configuration, with horizontal position indicating the signed SHAP contribution in kelvin and colour denoting the corresponding feature value from low (blue) to high (red). Features are ordered according to decreasing mean absolute SHAP value.*

To elucidate the physical basis of the temperature predictions, SHAP analysis was performed on the regression head of the PINN, as shown in Fig. 8. The feature ranking is dominated by near-icosahedral Voronoi motifs, demonstrating that local short-range atomic order provides the principal structural information used by the model to infer the thermodynamic state. Several highly coordinated motifs, including 〈0,2,8,2,0,0,0〉, 〈0,2,8,4,0,0,0〉, 〈0,2,8,1,0,0,0〉, and the full icosahedral 〈0,0,12,0,0,0,0〉 motifs, demonstrate significant SHAP contributions, where elevated motif populations are typically linked to negative impacts on the projected temperature. This suggests that larger populations in these locally ordered environments adjust the forecast toward cooler temperatures, aligning with their link to structurally stable glassy states. Cu composition ranks as one of the most significant characteristics and shows a wide range of both positive and negative effects, underscoring the combined impact of chemical makeup and local structural arrangement on the deduced thermal condition. Conversely, the logarithm of the quench rate shows a reduced overall SHAP magnitude yet a significant directional trend: increased quench rates tend to yield positive SHAP values, indicating higher predicted temperatures. This behavior aligns with maintaining elevated fictive temperatures during quick cooling. The relatively minor role of quench rate indicates that the thermal history is somewhat embedded implicitly in the resulting local structural patterns, whereas the explicit quench-rate variable offers additional processing details. In summary, the SHAP analysis reveals that the PINN prediction is primarily influenced by physically relevant local structural descriptors instead of functioning as a black box linking input features to temperature.

### 3.8 Sensitivity to physics-loss weighting

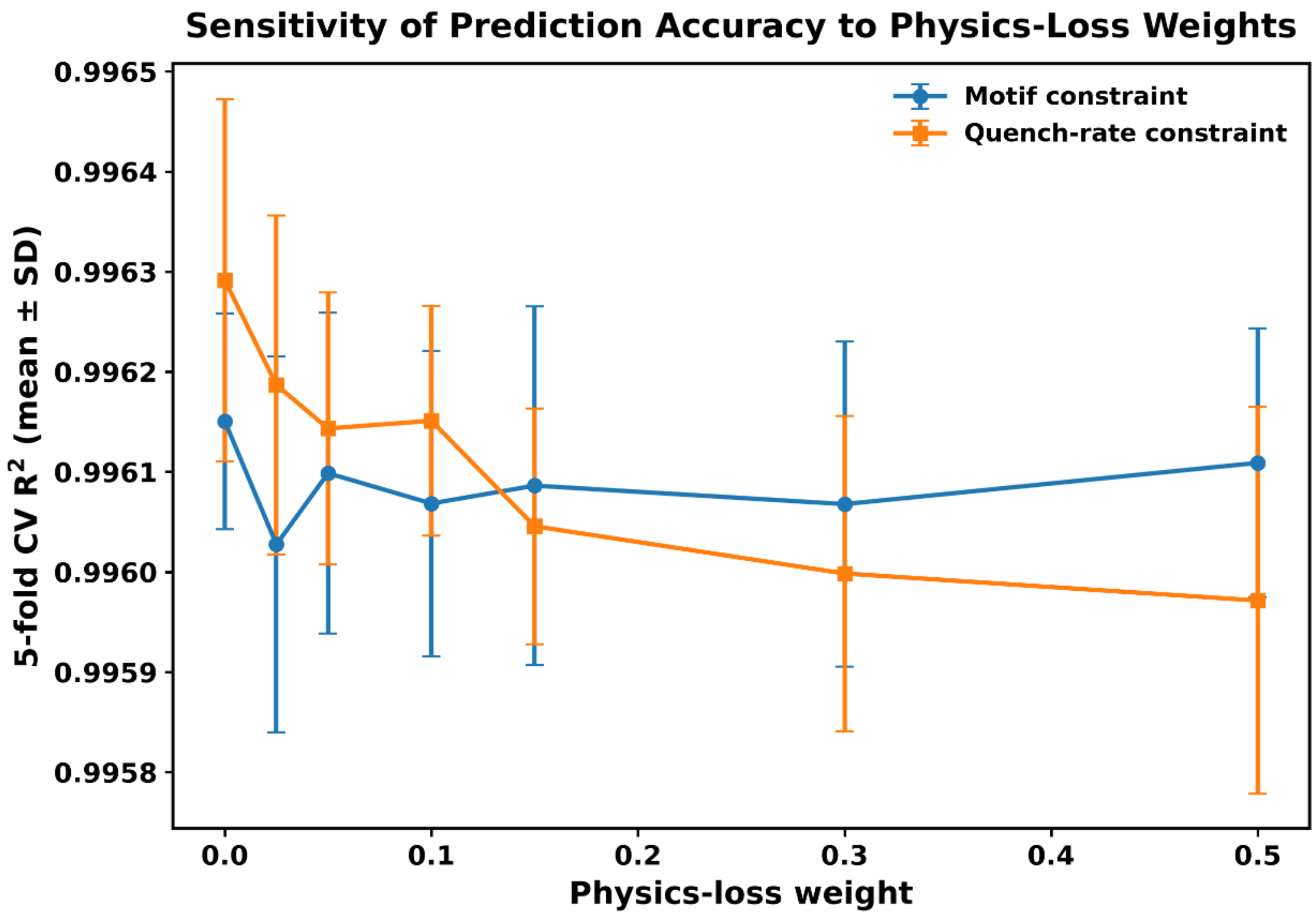

***Figure 9.*** *Heatmap showing the elemental compositions (at.%) of the ten optimized Co-based superalloys. The color scale represents the atomic percentage of each alloying element, highlighting common compositional trends and variations among the proposed alloy chemistries.*

The influence of the physics-loss weights on predictive performance was evaluated using five-fold trajectory-grouped cross-validation, as shown in Fig. 9. The mean cross-validation $R^2$ remains consistently high across the investigated weighting range, with values approximately between 0.9958 and 0.9965. For the motif constraint, the mean $R^2$varies only weakly around 0.9961 as the constraint weight increases from zero to 0.5, with no systematic loss of predictive performance. The quench-rate constraint exhibits a slightly stronger but still minor decrease in mean $R^2$, from approximately 0.9963 at zero weight to approximately 0.9960 at the largest investigated weight. The magnitude of these changes is small relative to the fold-to-fold variability, as indicated by the overlapping error bars.

These results demonstrate that the predictive accuracy of the PINN is largely insensitive to the weighting of the imposed physical constraints within the investigated range. Importantly, increasing the physics-loss contribution does not result in a substantial deterioration of the temperature prediction, indicating that the physical terms can be incorporated into the optimization objective without compromising the data-driven component of the model. The weak sensitivity also suggests that the observed physical consistency is not the result of selecting an excessively large constraint weight at the expense of predictive accuracy. Instead, the physics terms provide additional restrictions on the learned solution while retaining essentially the same regression performance.

### 3.9 Hyperparameter Selection: Regression Accuracy and Regime Classification

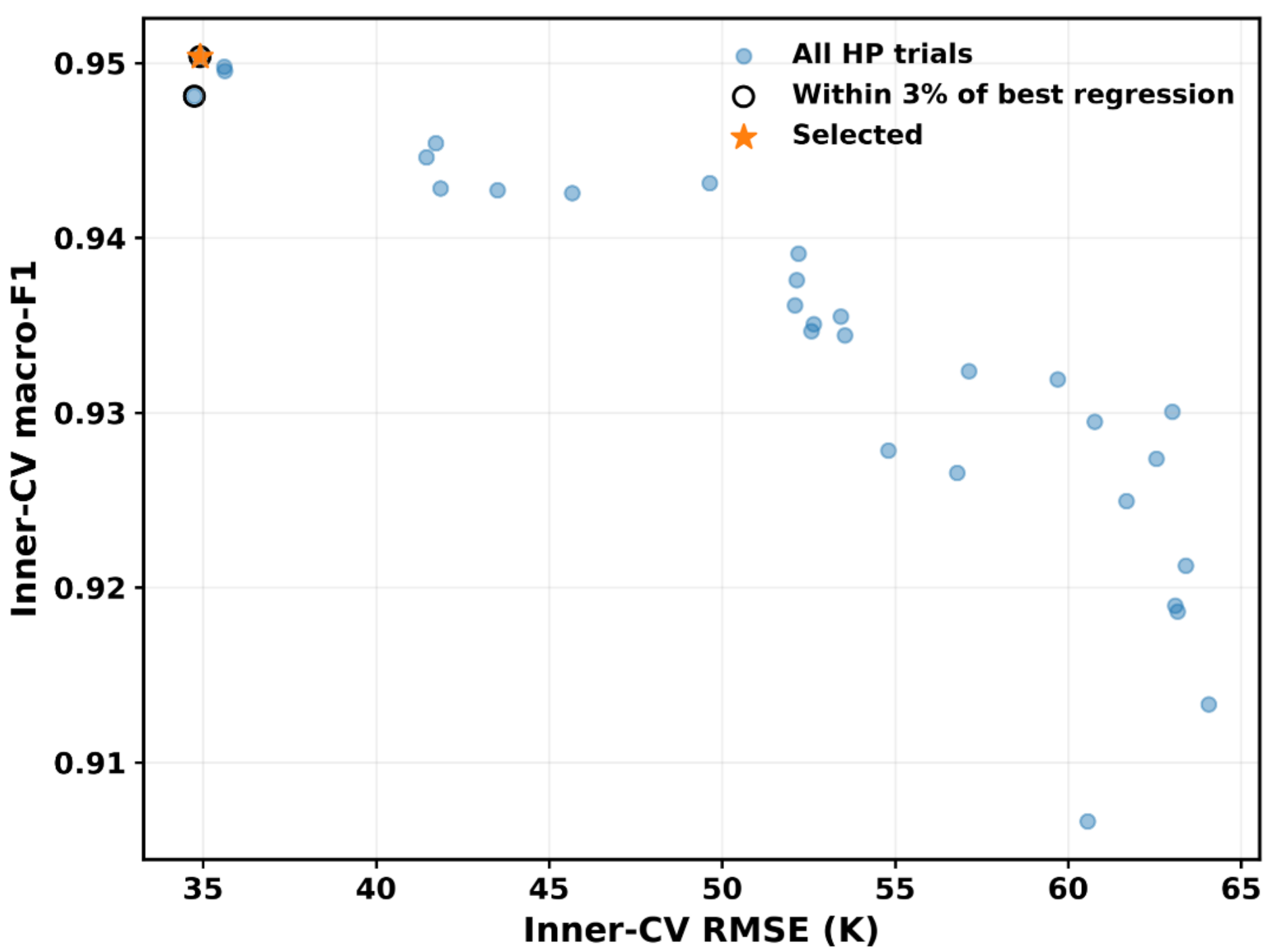

***Figure 10.*** *Hyperparameter-selection temperature-regression error and thermal-regime classification macro-F1. The selected configuration is constrained to remain within 3% of the best inner-CV regression MSE, after which macro-F1 is used as the secondary selection criterion. This prevents collapse of the auxiliary classifier without allowing classification performance to dominate the primary regression task.*

Figure 10 summarizes the inner cross-validation performance obtained for the explored hyperparameter configurations, with the inner-CV RMSE representing the regression objective and macro-F1 representing the auxiliary thermal-regime classification objective. The results show a clear concentration of high-performing configurations toward the upper-left region of the parameter space, where both low regression error and high classification performance are achieved. Configurations with an inner-CV RMSE of approximately 35–45 K generally maintained macro-F1 values above 0.94, whereas configurations with progressively larger regression errors exhibited a systematic reduction in classification performance. The selected configuration, indicated by the orange star, lies near the favorable joint-performance region, with an inner-CV RMSE of approximately 35 K and a macro-F1 of approximately 0.95. Notably, the selected solution does not achieve its regression performance at the expense of the auxiliary classification task, indicating that the chosen hyperparameter setting provides a balanced representation of the two learning objectives. This joint behavior supports the use of the selected configuration for the subsequent evaluation of the physics-informed model.

### 3.10 Benchmarking against conventional machine-learning models

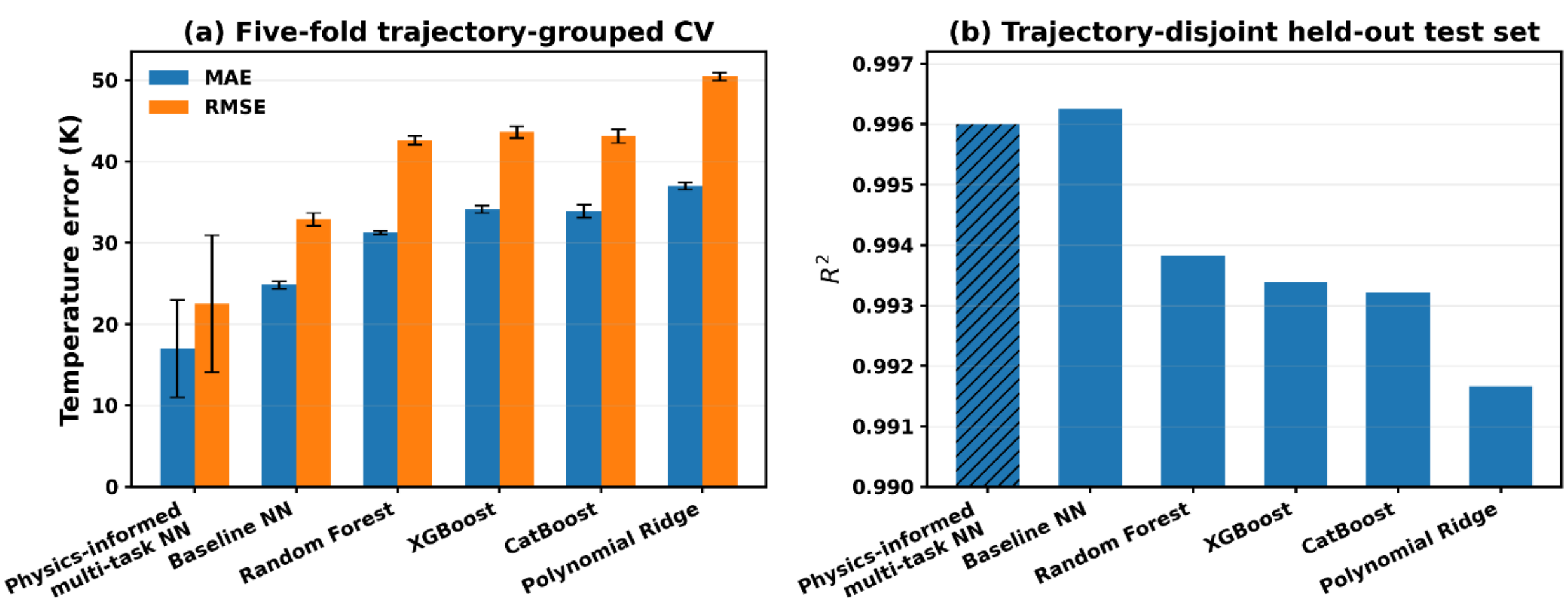

***Figure 11.*** *Comparison of the predictive performance of the proposed physics-informed multi-task neural network (PINN) with conventional machine-learning models. (a) Cross-validated regression performance, quantified using the root-mean-square error (RMSE), and (b) corresponding coefficient of determination ($R^2$) for the proposed PINN and baseline models, including Random Forest, Gradient Boosting, XGBoost, CatBoost, Support Vector Regression, and polynomial ridge regression. All models were evaluated using the same trajectory-disjoint data partition and preprocessing protocol.*

The predictive capability of the proposed physics-informed multi-task neural network was evaluated against a baseline neural network and established regression models using an identical trajectory-grouped cross-validation protocol. As shown in Fig. 11a, the proposed model exhibits the lowest temperature prediction error, with substantially smaller MAE and RMSE than the baseline neural network, Random Forest, XGBoost, CatBoost, and Polynomial Ridge models. The baseline neural network provides the closest conventional benchmark, whereas the tree-based and polynomial models exhibit progressively larger prediction errors. This trend is also reflected in the independent trajectory-disjoint test set (Fig. 11b), where the proposed model maintains a high coefficient of determination while achieving lower absolute prediction error than the competing models. The comparable $R^2$ values of the neural-network models indicate that all models capture the dominant temperature–structure relationship; however, the reduced MAE and RMSE demonstrate that incorporating the physically motivated constraints improves the accuracy of temperature estimation, particularly in terms of absolute prediction error. These results indicate that the benefit of the physics-informed formulation is not simply improved curve fitting, but a more accurate and physically constrained representation of the structure–thermodynamic-state relationship.

## 4. Conclusion

This study presents a physics-informed multi-task neural network framework designed to create an interpretable and physically coherent link between local atomic arrangement and thermodynamic conditions in Cu–Zr metallic glasses. The key findings can be outlined below:

- The suggested PINN showed significant predictive accuracy across the entire temperature range examined. In the independent trajectory-disjoint test set, the ensemble-mean model recorded an MAE of 23.44 K, an RMSE of 31.11 K, and an $R^2$ of 0.9965, with forecasts closely aligning with the optimal parity relationship from about 50 to 2000 K. Residual analysis indicated that errors mainly clustered near zero, with most predictions falling within roughly ±50 K, although deviations from Gaussian behavior were noted in the distribution tails. These findings show that the local Voronoi structural descriptors hold enough information to accurately reconstruct the thermodynamic state while ensuring generalization over previously unencountered molecular-dynamics trajectories.

- A significant finding of the research is that adding physical constraints did not necessitate compromising predictive accuracy. The near-icosahedral motif constraint and quench-rate causality constraint were integrated directly into the training objective using autograd-derived gradient residuals, thus promoting physically significant trends during model optimization instead of evaluating them solely post-training. Sensitivity analysis indicated that the regression performance was largely unaffected across the examined physics-loss weighting range, with five-fold cross-validation $R^2$ values consistently staying around 0.9958 to 0.9965. This shows that the applied physical constraints serve as valuable limitations on the learned outcome without significantly undermining its data-driven prediction ability.
- The auxiliary classification of thermal regimes further showed that the structural representation acquired by the shared network holds information regarding discrete thermodynamic states alongside continuous temperature. Misclassification mainly happened between neighboring regimes, with no direct mix-up between the glass and liquid categories, aligning with the seamless nature of the liquid-to-glass structural shift.
- The analysis of interpretability offers deeper physical understanding of the learned structure–state connection. Ensemble SHAP analysis showed that the prediction is primarily influenced by local Voronoi-motif descriptors instead of just composition. Specifically, multiple near-icosahedral motifs rank among the highest in feature importance, reinforcing the physical prior incorporated in the PINN formulation. The difference between motif prevalence and SHAP importance highlights that the most common structural environments aren't always the most useful for predicting temperature. Rather, the thermodynamic condition is collectively represented through a structure of interrelated local surroundings, with the near-icosahedral group standing out as a notably strong structural signature.
- Benchmarking against random forest, gradient-boosted-tree, and polynomial-ridge models under the same evaluation protocol showed that the proposed PINN achieved competitive regression accuracy while providing explicit physics-based constraints and simultaneous thermal-regime classification. This demonstrates that incorporating physical priors does not require sacrificing predictive performance relative to conventional machine-learning approaches.

The findings indicate that physical principles can be integrated directly into motif-level machine learning for metallic glasses without significantly reducing predictive accuracy, while also enhancing interpretability and reinforcing physically relevant structural and processing patterns. The integration of multi-task learning, differentiable physics constraints, trajectory-disjoint validation, ensemble-based uncertainty evaluation, and SHAP analysis offers a comprehensive framework for advancing from solely data-driven or post-hoc explainable models to physically constrained structure–property learning in disordered materials. While the current research focuses on the composition and thermal-processing aspects indicated by the existing Cu–Zr molecular-dynamics dataset, the framework establishes a basis for future expansion to wider alloy compositions, experimentally obtained structural descriptors, and inverse-design challenges that necessitate physically consistent forecasts beyond the original training range.

**Conflict of Interest Statement**

The authors declare that they have no known competing financial interests or personal relationships that could have influenced the work reported in this paper.

**Credit Authorship**

Prashil S. Joshi: Conceptualization, Methodology, Visualization, Investigation, Writing – original draft, reviewing, Resources.

**Data Availability Statement**

The primary dataset used to train and validate the PINN model in this study was compiled from the dataset in [32], publicly available at https://github.com/branicio/cu-zr-disorder-motifs. Additional data supporting the findings of this study can be made available upon reasonable request.

**References:**

[1] M. Ashby, A. Greer, Metallic glasses as structural materials, Scripta Materialia 54 (2006) 321–326. https://doi.org/10.1016/j.scriptamat.2005.09.051.

[2] J. Schroers, W.L. Johnson, Ductile Bulk Metallic Glass, Phys. Rev. Lett. 93 (2004) 255506. https://doi.org/10.1103/PhysRevLett.93.255506.

[3] H.W. Sheng, W.K. Luo, F.M. Alamgir, J.M. Bai, E. Ma, Atomic packing and short-to-medium-range order in metallic glasses, Nature 439 (2006) 419–425. https://doi.org/10.1038/nature04421.

[4] Y.Q. Cheng, E. Ma, Atomic-level structure and structure–property relationship in metallic glasses, Progress in Materials Science 56 (2011) 379–473. https://doi.org/10.1016/j.pmatsci.2010.12.002.

[5] W.H. Wang, Dynamic relaxations and relaxation-property relationships in metallic glasses, Progress in Materials Science 106 (2019) 100561. https://doi.org/10.1016/j.pmatsci.2019.03.006.

[6] J. Ding, Y.-Q. Cheng, H. Sheng, M. Asta, R.O. Ritchie, E. Ma, Universal structural parameter to quantitatively predict metallic glass properties, Nat Commun 7 (2016) 13733. https://doi.org/10.1038/ncomms13733.

[7] F. Zhu, S. Song, K.M. Reddy, A. Hirata, M. Chen, Spatial heterogeneity as the structure feature for structure–property relationship of metallic glasses, Nat Commun 9 (2018) 3965. https://doi.org/10.1038/s41467-018-06476-8.

[8] H.-B. Yu, W.-H. Wang, K. Samwer, The β relaxation in metallic glasses: an overview, Materials Today 16 (2013) 183–191. https://doi.org/10.1016/j.mattod.2013.05.002.

[9] Y.Q. Cheng, H.W. Sheng, E. Ma, Relationship between structure, dynamics, and mechanical properties in metallic glass-forming alloys, Phys. Rev. B 78 (2008) 014207. https://doi.org/10.1103/PhysRevB.78.014207.

[10] W.P. Weeks, K.M. Flores, Structural building-blocks of disordered Cu-Zr alloys, Acta Materialia 265 (2024) 119624. https://doi.org/10.1016/j.actamat.2023.119624.

[11] Z.W. Wu, F.X. Li, C.W. Huo, M.Z. Li, W.H. Wang, K.X. Liu, Critical scaling of icosahedral medium-range order in CuZr metallic glass-forming liquids, Sci Rep 6 (2016) 35967. https://doi.org/10.1038/srep35967.

[12] R. Soklaski, Z. Nussinov, Z. Markow, K.F. Kelton, L. Yang, Connectivity of icosahedral network and a dramatically growing static length scale in Cu-Zr binary metallic glasses, Phys. Rev. B 87 (2013) 184203. https://doi.org/10.1103/PhysRevB.87.184203.

[13] J. Ding, Y.-Q. Cheng, E. Ma, Full icosahedra dominate local order in Cu64Zr34 metallic glass and supercooled liquid, Acta Materialia 69 (2014) 343–354. https://doi.org/10.1016/j.actamat.2014.02.005.

[14] N. Mattern, P. Jóvári, I. Kaban, S. Gruner, A. Elsner, V. Kokotin, H. Franz, B. Beuneu, J. Eckert, Short-range order of Cu–Zr metallic glasses, Journal of Alloys and Compounds 485 (2009) 163–169. https://doi.org/10.1016/j.jallcom.2009.05.111.

[15] Y.Q. Cheng, E. Ma, H.W. Sheng, Atomic Level Structure in Multicomponent Bulk Metallic Glass, Phys. Rev. Lett. 102 (2009) 245501. https://doi.org/10.1103/PhysRevLett.102.245501.

[16] A. Hirata, P. Guan, T. Fujita, Y. Hirotsu, A. Inoue, A.R. Yavari, T. Sakurai, M. Chen, Direct observation of local atomic order in a metallic glass, Nature Mater 10 (2011) 28–33. https://doi.org/10.1038/nmat2897.

[17] S.Y. Wang, C.Z. Wang, M.Z. Li, L. Huang, R.T. Ott, M.J. Kramer, D.J. Sordelet, K.M. Ho, Short- and medium-range order in a Zr 73 Pt 27 glass: Experimental and simulation studies, Phys. Rev. B 78 (2008) 184204. https://doi.org/10.1103/PhysRevB.78.184204.

[18] J. Ding, S. Patinet, M.L. Falk, Y. Cheng, E. Ma, Soft spots and their structural signature in a metallic glass, Proc. Natl. Acad. Sci. U.S.A. 111 (2014) 14052–14056. https://doi.org/10.1073/pnas.1412095111.

[19] Z.-Y. Yang, Y.-J. Wang, Ergodic Structural Diversity Predicts Dynamics in Amorphous Materials, Front. Mater. 9 (2022) 855681. https://doi.org/10.3389/fmats.2022.855681.

[20] S. Yuan, A. Liang, C. Liu, A. Nakano, K. Nomura, P.S. Branicio, Uncovering metallic glasses hidden vacancy-like motifs using machine learning, Materials & Design 233 (2023) 112185. https://doi.org/10.1016/j.matdes.2023.112185.

[21] E.J. Gurniak, S. Yuan, X. Ren, P.S. Branicio, Harnessing graph convolutional neural networks for identification of glassy states in metallic glasses, Computational Materials Science 244 (2024) 113257. https://doi.org/10.1016/j.commatsci.2024.113257.

[22] N. Leimeroth, J. Rohrer, K. Albe, General purpose potential for glassy and crystalline phases of Cu-Zr alloys based on the ACE formalism, Phys. Rev. Materials 8 (2024) 043602. https://doi.org/10.1103/PhysRevMaterials.8.043602.

[23] D. Wei, J. Yang, M.-Q. Jiang, B.-C. Wei, Y.-J. Wang, L.-H. Dai, Revisiting the structure–property relationship of metallic glasses: Common spatial correlation revealed as a hidden rule, Phys. Rev. B 99 (2019) 014115. https://doi.org/10.1103/PhysRevB.99.014115.

[24] G. Liu, S. Sohn, S.A. Kube, A. Raj, A. Mertz, A. Nawano, A. Gilbert, M.D. Shattuck, C.S. O'Hern, J. Schroers, Machine learning versus human learning in predicting glass-forming

ability of metallic glasses, Acta Materialia 243 (2023) 118497. https://doi.org/10.1016/j.actamat.2022.118497.

[25] S. Lundberg, S.-I. Lee, A Unified Approach to Interpreting Model Predictions, (2017). https://doi.org/10.48550/ARXIV.1705.07874.

[26] H. Gleiter, Nanoglasses: a new kind of noncrystalline materials, Beilstein J. Nanotechnol. 4 (2013) 517–533. https://doi.org/10.3762/bjnano.4.61.

[27] S.H. Nandam, Y. Ivanisenko, R. Schwaiger, Z. Śniadecki, X. Mu, D. Wang, R. Chellali, T. Boll, A. Kilmametov, T. Bergfeldt, H. Gleiter, H. Hahn, Cu-Zr nanoglasses: Atomic structure, thermal stability and indentation properties, Acta Materialia 136 (2017) 181–189. https://doi.org/10.1016/j.actamat.2017.07.001.

[28] D. Şopu, K. Albe, J. Eckert, Metallic glass nanolaminates with shape memory alloys, Acta Materialia 159 (2018) 344–351. https://doi.org/10.1016/j.actamat.2018.08.034.

[29] S. Adibi, Z.-D. Sha, P.S. Branicio, S.P. Joshi, Z.-S. Liu, Y.-W. Zhang, A transition from localized shear banding to homogeneous superplastic flow in nanoglass, Applied Physics Letters 103 (2013) 211905. https://doi.org/10.1063/1.4833018.

[30] P.S. Joshi, X. Ren, P.S. Branicio, Tailoring Cu-Zr gradient nanoglass structures: Influence of nanoparticle size and cooling rates on glass-glass interfaces, Intermetallics 178 (2025) 108633. https://doi.org/10.1016/j.intermet.2024.108633.

[31] M. Raissi, P. Perdikaris, G.E. Karniadakis, Physics-informed neural networks: A deep learning framework for solving forward and inverse problems involving nonlinear partial differential equations, Journal of Computational Physics 378 (2019) 686–707. https://doi.org/10.1016/j.jcp.2018.10.045.

[32] J. Wang, S. Yuan, P.S. Branicio, Explainable machine learning reveals that local structural motifs encode the thermodynamic state across the CuZr metallic glass-forming range, Computational Materials Science 272 (2026) 114836. https://doi.org/10.1016/j.commatsci.2026.114836.